# Fast and direct visualization of piezo-generated charges at the nanoscale using direct piezoelectric force microscopy

A. Gomez[1]*,

[1]Instituto de Ciencia de Materiales de Barcelona (ICMAB-CSIC), Campus UAB, 08193, Bellaterra, España

*Corresponding author: agomez@icmab.es

**ABSTRACT**

The denominated "surface charge scraping" mechanism was discovered in 2014 by using a new Atomic Force Microscopy (AFM) based mode called Charge Gradient Microscopy. The measurements to probe such mechanism are achieved with the use of a current-to-voltage converter: a transimpedance amplifier (TIA). However, the use of an incorrect approximation, named Gain BandWidth Product (GBWP) to calculate TIA's BandWidth (BW) could mislead to an incorrect data interpretation. By measuring at higher frequencies than permitted, the amplifier is used as a current-to-voltage converter, in conditions where it behaves as a charge-to-voltage converter. In this manuscript, we report the specific conditions in which the transfer function of the same electronic circuit topology is valid, while we spot both ringing and unstable amplifiers artifacts in the published data. We finally perform physical measurements in similar conditions as reported, but fully respecting the BandWidth (BW) of the system. We find that the charge collected is way below the values reported in such publication, diminishing or even nullifying the impact of a possible charge scrapping mechanism. These findings pave the way to employ Direct Piezoelectric Force Microscopy (DPFM) as a fast ferroelectric nanoscale characterization tool.

**Significance Statement**
Denying a novel physical phenomena is as important as reporting it. This is why we present this manuscript in which we deny the new physical phenomena presented in the PNAS manuscript "Charge Gradient Microscopy" with the name of "Surface Charge Scraping Mechanism". Our manuscript analyzes why this new phenomena is reported using a current-to-voltage converter which was incorrectly used, outside the specifications in which it can be employed. In these conditions, we prove the signal recorded cannot be related to the input current value. Moreover, a complete new set of measurements is included in the manuscript providing a new window into the nanoscale and fast ferroelectric characterization techniques based in Atomic Force Microscope.

**Introduction**
Current-to-Voltage converters are among the most used instruments at scientific laboratories for many different applications. For instance, they are used to characterize solar cells(1, 2), analyze material's electrical properties(3), performing transport measurements(4), STM and AFM microscopes(5, 6), superconductors characterization (7, 8) or charge quantification (9), between many others. The industrial applications comprises the use in transducers(10), Ethernet connections(11), fiber optic communications(12), smart phone touch screen(13) or Internet-Of-things nodes(14). The rising need of highly precision, low-powered current measurement circuits for the up-coming applications is spreading the research of novel architectures. With such enormous implications, the understanding of this kind of electronic equipment is crucial for scientists outside of instrumentation and electronic engineering to exploit its capabilities.

The family of current-to-voltage converters is typically called Trans-Impedance Amplifiers (TIA), from now on. In TIAs, a resistor is connected between the inverting-input and the output pins of the amplifier(15, 16). The feedback path is formed by two elements, a feedback resistor, $R_f$, and a parallel capacitor, $C_f$, see **Figure 1a**. Even if the circuit does not possess a capacitor physically, a strain capacitance, inherent from the circuit, has to be considered(17). In this scheme, the "unknown" current, the physical quantity to be measured, is feed throughout the inverting-input of the amplifier. The amplifier tries to compensate such input current, by providing a current flow from the output to the inverting input, of the same magnitude, but in the opposite direction, $I_f$ of the **Figure 1a**. In this scheme a suitable input resistance at the inverting path should be introduced here as to denote the resistance inherent to input wiring. For simplification purposes, no strain capacitor is considered in the input resistance, however a more complex analysis of this circuit can be found in literature(18). The TIA circuit topology converts current into voltage with the following gain:

$$V_{out} = -R_f I_{in} \quad (1)$$

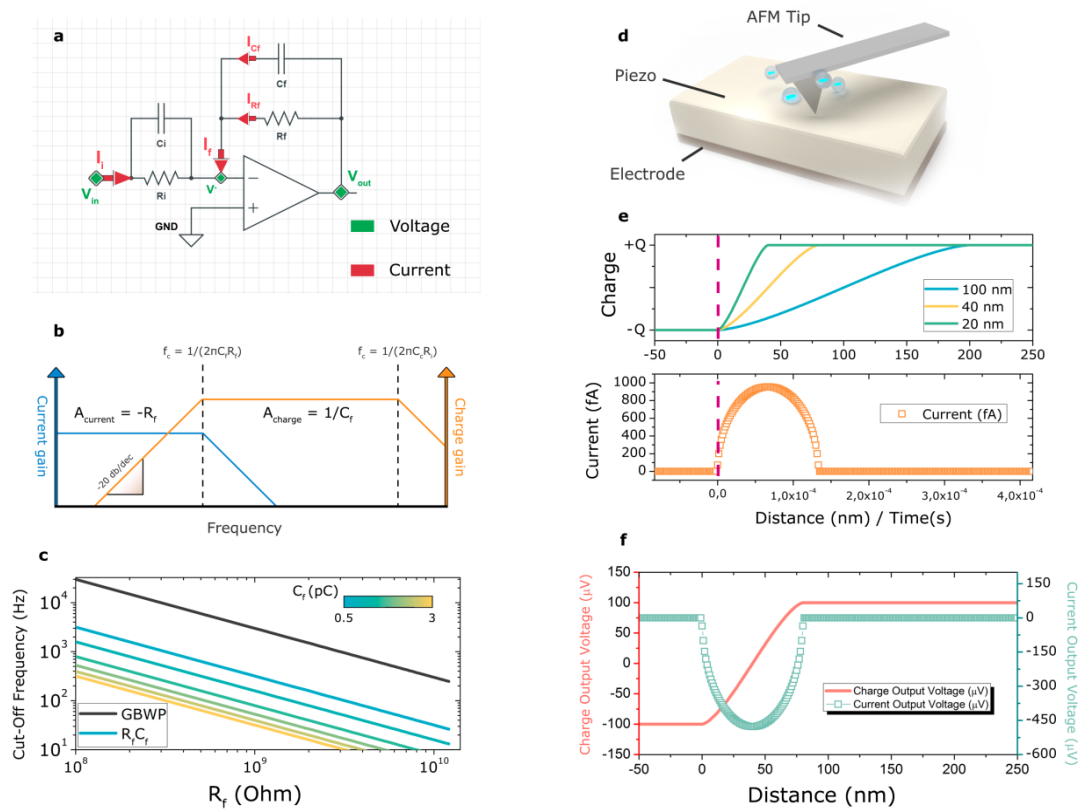

**Figure 1**: A TransImpedance Amplifier (TIA) analysis, **a**, circuit topology describing the transimpedance amplifier configuration. **b**, Current gain and Charge gain as a function of the frequency; **c**, Cut-off frequency vs the feedback resistor value, comparing the use of Gain-Bandwidth-Product approximation and the $R_fC_f$ approximation, the colors denote different values of the feedback capacitor; **d** 3D physical model of the AFM tip scanning a piezoelectric charge generation, **e**, Model of the piezo-generated charge obtained when the tip crosses from a down polarization domain to an up polarization state, Up and the derivative of the generated charge, down, for the case of a 40 nm tip-sample contact radius. **f**, Voltage output of a charge amplifier, red line, and a TIA, green squares, obtained when the tip crosses two antiparallel domains.

Despite the apparent simplicity of this equation, the exact same topology can act as a voltage-to-voltage converter and as a charge-to-voltage converter, increasing its complexity. Specifically, the exact same circuit layout may behave differently depending into the measurement conditions. In the case of a TIA, the amplifier behaves as a current-to-voltage converter, however, in other circumstance; the exact same circuit could act as a charge-to-voltage converter. And, to add more complexity, in other conditions the amplifier may behave as a voltage-to-voltage converter, an inverting amplifier(15). Hence it is of crucial importance to employ the adequate transfer function at the right conditions. Between all the measurement conditions, the frequency of the input signal is, likely, the most important parameter. If the input frequency is below the low pass band filter formed by $R_f$ and $C_f$, the circuit behaves as a TIA(19). However, if the frequency increases, the current gain of the topology decreases with a slope of -20 db/dec and more importantly, a charge-to-voltage gain appears, the amplifier begins to amplify charge rather than current, see bode plot of **Figure 1b**. This gain is of special importance while measuring charge generators. Underneath such bode plot there is a physical explanation. For low frequencies, the feedback capacitor can be considered as an open circuit, hence, almost no current flows through it. Consequently, if the current is primarily low frequency, the current will circulate through the feedback resistor. However, the impedance of a capacitor, $(Z_L = 1/{C_F \omega})$, being $\omega$ the frequency, decreases for higher frequency values. Hence, at higher speed, some charges may start to flow throughout the capacitor path, as the capacitor impedance could reach the value of the resistor impedance. Consequently, as frequency increases, the amplifier may stop amplifying current and start amplifying charge(20–24). At very high frequencies, the charge gain value given by G = 1/$C_f$ decreases with -20db/dec, see **Figure 1b**. We performed a summary between the gains, BW and signals to measure for each amplifier, which is depicted in **table 1**.

| | Inverting amplifier topologies | | |
|---|---|---|---|
| | **Converter type** | **Simplified Transfer Function** | **Frequency range** |
| TIA | Current-to-Voltage | $V_{out} = -R_f I_{in}$ | $0 < f < 1/2\pi C_f R_f$ |
| Charge Amplifier | Charge-to-Voltage | $V_{out} = Q_{in}/C_f$ | $1/2\pi C_f R_f < f < 1/2\pi C_p R_i$ |
| Inverting Amplifier | Voltage-to-Voltage | $V_{out} = -R_f/R_i V_{in}$ | $0 < f < (GBWP)R_i/R_f$ |

**Table 1**: Comparison between the different converters sharing the same inverting amplifier topology, for the case of converter type, transfer function and frequency range.

As equally important as selecting the correct transfer function, calculating the Bandwidth (BW) with the right approximation becomes mandatory. Common errors rely into applying a BW calculation procedure for an incorrect circuit topology(25). A comparison between using Gain-Bandwidth-Product (GBWP) approximation and RC filter approximation is depicted **Figure 1c**. For the case of a high $R_f$ of 500M, the GBWP gives substantial higher bandwidths compared to the case of using the RC low-pass filter approximation. Such increase attain up to two orders of magnitude, obtained while comparing GBWP and RC methods, considering a favorable low value of feedback capacitor, 1pC. If the GBWP is used as an approximation for both charge and transimpedance amplifiers, the BW values obtained are not correct, and hence, the behavior of the amplifier topology cannot be predicted and accurately modeled.

**Experimental part**

A special case arises for a new physical phenomena reported by Hong et al. in 2014 relating ferroelectric materials(26). To report such phenomena, a nanometer size metallic needle is employed scans a ferroelectric surface, with a TIA attached to the needle. The tip is employed with the intention of scanning different ferroelectric domain structures while measuring the current collected by the tip, simultaneously. From a technical point of view, CGM has similarities with the mode Direct Piezoelectric Force Microscopy (DPFM). Though, in DPFM, the measurements were carried out at a radical 1000x slower speed while applying 100x times more force to the material. More importantly, an ultra-low leakage input bias amplifier is used with a very high (1TeraOhm) feedback resistor path. Such conditions were selected, specifically, to avoid the surface charge scrapping effect reported(27). While performing our measurements, we realize that, indeed, the piezoelectric effect overcomes the new surface charge scraping phenomena which, in principle, contradicts Hong et al manuscript(26). In order to explain why this occurs, we consider an antiparallel out of plane domain configuration to depict the charge generated by piezoelectric effect, see **Figure 1e**. If we now describe how the AFM tip crosses two consecutive antiparallel ferroelectric domains, we obtain a graph depicting the generated charge profile. This charge is constant in the regions of the same domain structure, however when the tip crosses antiparallel domains, the charge changes dramatically, see **Figure 1e**. The process can be described considering a tip-sample contact area of a circle, in which the different circle radiuses, provides a more localized or disperse charge gradient. To provide a clearer view, different tip-sample contact area radius of 100, 40 and 20 nm were calculated; see **Figure 1e**, in which each color represents different tip-sample contact areas. We see that the absolute charge displacement is not dependent upon tip-sample contact area, as expected for piezoelectric effect, which only depends upon the force applied(28). However, the profile shape obtained depends upon the contact area and scan speed: the smaller the contact area, the higher the current recorded and the faster the input signal. At this point, we can numerically differentiate the charge profile to obtain the generated current. For simplicity, we choose the charge profile corresponding to the 40 nm tip radius, see **Figure 1e,** which shows the graph of the generated current. To obtain this curve, a Q of 0.1fC is assuming while a scanning speed of 600 µm/s is considered. Through this simple model, we are able to theoretically calculate the amplifier output voltages, for the case of a TIA and charge amplifier, see **Figure 1f**. The voltage profiles are obtained with $R_f$ of 500MOhm, $C_f$ of 0.5pC and at the -3db roll off frequency, calculated with the RC low pass filter formula. We can see that, in the best case scenario, the output voltage for charge gain is comparable to the output voltage for the TIA gain.

For the case of CGM, the bandwidth of a TIA is calculated as an inverting amplifier, using the GBWP approximation(26); the amplifier gain corresponds to a TIA, while the frequency in which the amplifier is used belongs to the charge amplifier range. This mix of three different parameters for the same circuit topology is causing, what we humbly believe, is an incorrect data interpretation. In order to explain their results, we firstly replicate a part of the experiments in a TIA with a much higher feedback resistor. A value of 10GOhm was selected for $R_f$ which effectively increases the gain of the TIA, however at a cost of lowering the bandwidth of the system(29). Through this approach, we scanned in DPFM mode a test sample, which comprises a Periodically Poled Lithium Niobate (PPLN)(30), by applying different amounts of force values-see **Figure 2a**.

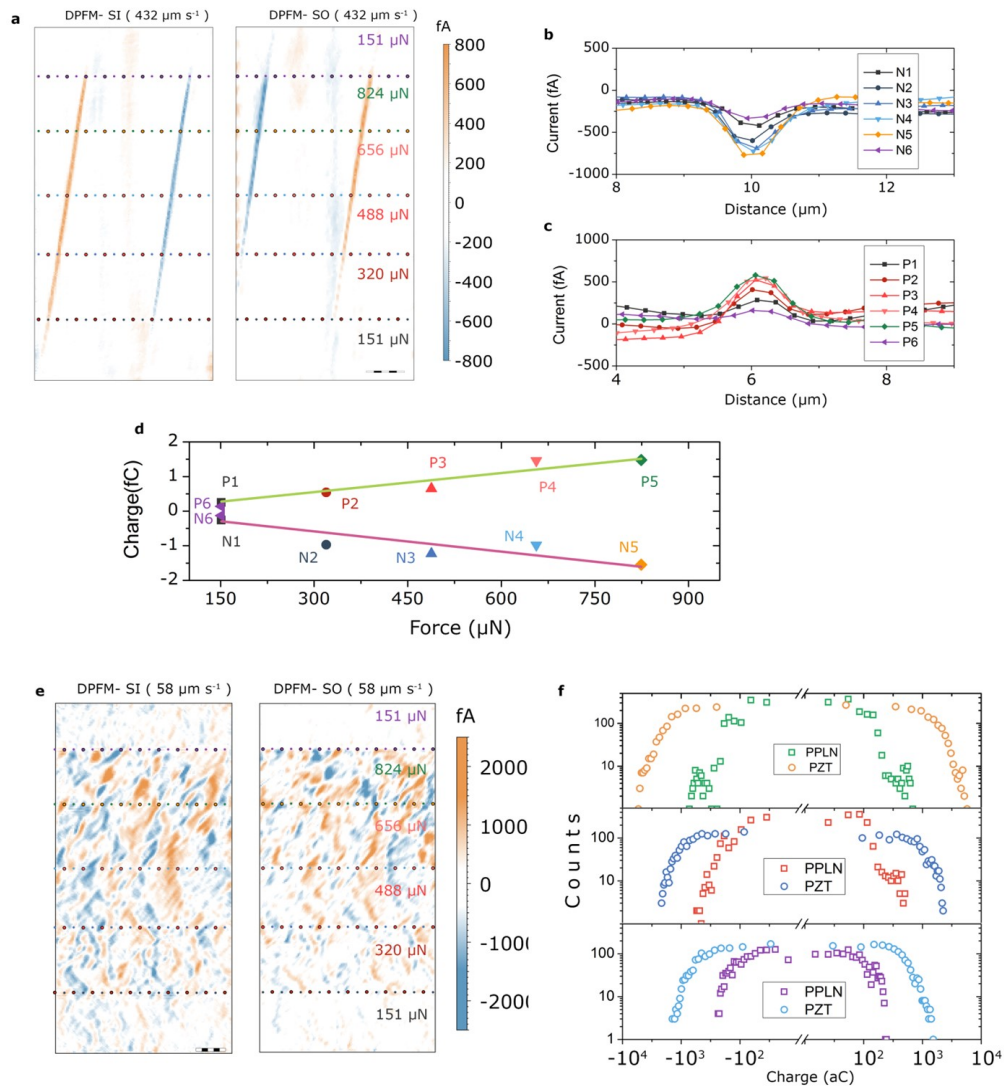

**Figure 2**: DPFM experiments carry out in CGM speed conditions. **a**, DPFM-Si and DPFM-So images obtained for a Periodically Poled Lithium Niobate (PPLN) test sample in which different loads were applied while performed a bottom-to-top scan. **b** and **c**, extracted profiles for the case of DPFM-Si and DPFM-So, respectively, at different applied loads. **d**, relationship between the applied charge and applied Force, obtained by multiplying the current recorded from the amplifier by the specific pixel time constant. **e**, DPFM-Si and DPFM-SO images obtained for the case of a PZT sample with natural domains grown by thermal treatment. f, Histograms obtained from the PPLN and PZT DPFM images, in which the current is multiply by the specific pixel time rate to depict the level of charge obtained, for different applied loads.

To obtain the data, rather than scanning the sample at very low speed, we employed a tip speed 432 µm/s, a similar value compared to CGM. While performing the image, which starts from bottom to top, we changed the load applied to the material, see dot horizontal lines delimiting each applied force in **Figure 2a**. From such image, it can be seen that the current recorded is proportional to the applied force. For each of the forces applied, we specifically extracted different current profiles obtained from the aforementioned images. We named the profiles as letter for "N" denoting the negative current values of DPFM-Si and "P" for the positive values of the DPFM-So image. The subsequent number denotes the specific force applied to the sample, meaning that 1 corresponds to the value of 151 µN, 2 = 320 µN, 3 = 488 µN... and so on. The data shows that both maximum value and area increases with an increase of the force applied. More importantly, we were able to integrate the current profile, to acquire the generated charge versus applied load-see **Figure 2d**. We were able to obtain a linear relationship between the charge and force, with $R^2$ of 0.93, as expected for piezoelectric effect. It is important to note that the speed 432 µm/s is very close to the values reported in CGM scans(26). To obtain the tip speed, we included the X-axis over-scan which is present in the vast majority of microscopes. To elucidate if charge scrapping is also important in other materials, we performed another set of measurements into a Lead Zirconate Titanate (PZT) material, which has a lower surface charge density compared with PPLN(31, 32). However, PZT $d_{33}$ piezoelectric constant is considerable larger than lithium niobate(30, 33). Hence, at this point, if surface charge scraping occurs, the charge obtained should be much lower than the case of PPLN. Our results, see **Figure 2e**, goes all the way around. For PZT the level of charge recorded is much higher than the case of PPLN. These results could not be explained if the charge scrapping phenomena is the predominant cause of the current recorded. In order to compare the results obtained in **Figure 2a** and **Figure 2e**, we calculated the charge histograms for each of the images. The charge is calculated by multiplying the specific pixel time constant, τ, times the current obtained(27). With these values, we obtained charge histograms which are plotted in **Figure 2f**, for different applied forces. Our results show that a higher force generates a higher charge amount. More importantly, we see the collected charge for PZT material is one order of magnitude higher than lithium niobate, independently of similar forces used. This result confirms that piezoelectricity is the predominant effect rather than the new phenomena of charge scrapping mechanism.

In order to explain why these measurements can be carried out, we focus our attention into CGM collected data. As denoted by **Table 1**, it is crucial to determine the behavior of the amplifier: current-to-voltage, voltage-to-voltage or charge-to-voltage amplifier. For the case of CGM, the tip speed employed to scan the material reach up to 70Hz, in images of 256 points, would output a data throughput of 256*70 = 17,9 kHz. The tip radius value of 40 nm used in the manuscript, is way smaller than the pixel size value, hence the tip diameter would not represent a further bandwidth limitation. This value is way too high for the case of an amplifier behaving like a TIA, and hence, the amplifier behaves as a charge amplifier with a completely different transfer function. In order to depict this idea, we have recreated the bode plot of a transimpedance amplifier with similar parameters as in the case of CGM, with $R_f$ = 500 MOhm and $C_f$ = 0.5 pC(25), see **Figure 3a**. Even with a very optimistic value for the feedback capacitor, none of the CGM frames published in PNAS complies with the necessary BW limitations imposed by the amplifier. We plotted, as different vertical lines the different speed rates used to demonstrate this situation. The vertical lines denotes the necessary bandwidth to acquire a reliable current measurements, obtained, by multiplying 256 px times the specific scan rate, 5 Hz, for red vertical line, 10 Hz for orange line, 20 Hz for yellow line, 30 Hz for green line and 70 Hz for blue line.

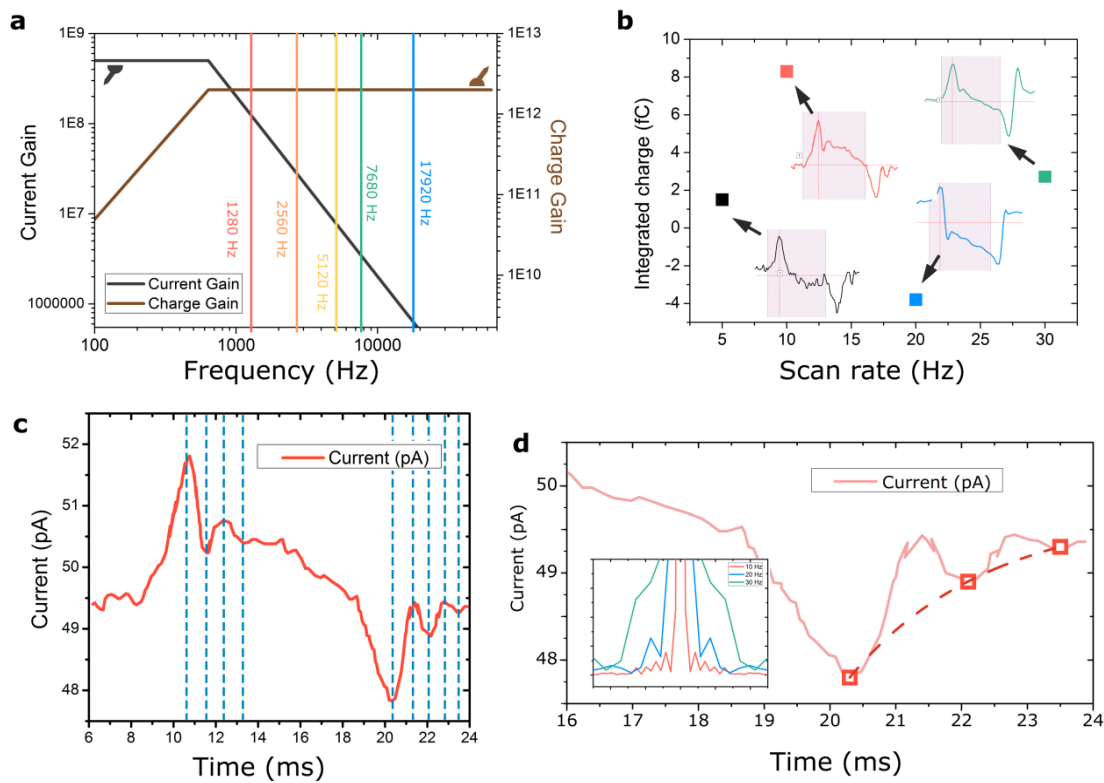

**Figure 3**: Analysis of the profiles obtained in CGM manuscript. **a**. Bode plot depicting the evolution of the TIA's and charge amplifier gains, versus the frequency. **b**. Integrated charge obtained for the different current profiles. **c**. Current profile for the case of 10 Hz scan, in which the maximum and minimum values of the signal are measured. **d**. 10Hz scan profile obtained from the manuscript with exponential fitting to find the time constant of the signal; the insert is the FFT of all the profiles, showing a similar behavior than a step-like response of a slow feedback amplifier.

At this point, we can conclude that the conditions necessary for a TIA are not met. Hence, the output voltage of the circuit is not related to the input current uniquely. To prove this situation from a physical point of view, we integrated all the profiles present in the supporting information, for the different scan rates of 5, 10, 20 and 30 Hz, the profiles were acquired with WebPlotDigitizer software, directly from the published figures. The results are plotted in **Figure 3b**. The results of the integrated charge are erratic with increasing scan rate, which can be a consequence of the unstable behavior of the amplifier and the wrong transfer function used. Each of the profiles is integrated through a full domain, obtaining the measured charge, the integral range is depicted as faded red background color. Such result is confusing, as the authors clearly show that the peak current increases with the increase scan rate, however the charge collected does not follow the same tendency. More importantly, the behavior of too much delay in the path of the amplifier could rise to different unwanted effects, being the "ringing" effect one of the most common(34). This effect is perfectly described in electronic literature(35–37). As we proved that the measurement speed is too high for the amplifier used, some of these artifacts should occur. To depict this situation, we specifically analyze the current profiles, selecting the 10 Hz profile, as it is the one that was integrated in the PNAS article(26). The profile is depicted in **Figure 3c**, as a read line, in which the vertical dotted lines denote the position of maximum and minimum of the ringing oscillation. As a consequence of a delay in the feedback of amplifiers a ringing effect typically occurs. This ringing effect present in step-like input functions has a constant frequency oscillation which depends mainly into the feedback amplifier path. We selected each of the minimum and maximum points of the profile in **Figure 3c** and calculated the specific half-period of the oscillation, which gives a value of 0,8 ± 0,2 ms . In the PNAS article, this ringing effect is interpreted as describing a physical phenomenon, related to different processes occurring inside the scrapping charge mechanism with the AFM tip. More importantly, the exponential decay function of the step like ringing response is calculated, see **Figure 3e**. The time constant found is 2,1 ms, which is in accordance with possible ringing effects as it is close to the RC time constant exponential decay presented in RC circuits. For piezoelectric transducers, this signal oscillation is often called "pop-corn effect" which is a known effect that occurs while using charge amplifiers(34). All of these possibilities are valid to explain the reason of this oscillation effect. This exact same frequency oscillation is interpreted as a new physical phenomena, without taking into account the possibility of being a ringing effect.

At this point, we have proved that the BW calculation used is incorrect, that piezoelectricity overcomes surface charge scrapping, that the measuring speed in CGM scans are not achievable by a TIA and more importantly, we have spotted ringing and oscillation effects as a consequence of a slow feedback amplifier path. Even with this clear and concise information, we want to emphasize the physical contradictions that can be found in the text. The only experiment performed to corroborate that the signal comes from surface charge screening, is performed with a lithium niobate sample which was heated up to a point that there is no surface charges. However, to acquire this image, a force of 30 nN is applied, while in the main text, it is written that a force of ~1 µN is needed to remove all the surface charge screening. Another important aspect is the surface recharge time needed for the process to occur. A video showing the AFM scanning multiple frames of the LTO sample, is available in the supporting information, in which in a 19 seconds duration, they performed 3 consecutive scans, saving both the trace and retrace images, up to a total of 6 frames(26). This means that the surface charge is scrapped, collected, and re-screened in time scales of 1- 2 seconds. This is contradictory with current literature examples, in which this rescreening process is well studied(38–40). In several other manuscripts, the surface screening recharge process occurs in times of minutes and even hours in some cases. The fact that the screening charge is recharged in time scales of seconds is contradictory with other manuscripts available. Even though that the charge measured is not accurate, the fact that the authors can see the ferroelectric domains in the images is of great interested for the community of ferroelectrics, in which there was only a mode available, Piezoresponse Force Microscopy(41–43). However such mode cannot be currently used as a single test to prove ferroelectricity(44), even though the efforts from the scientific community are close to this achievement(45–47).

**Conclusions**

In this manuscript we report the conditions in which an inverting amplifier acts as a current to voltage converter, a voltage to voltage converted or a charge to voltage converter. An intuitive view into the transfer functions available for TIA, inverting amplifier and charge amplifier is provided. We summarize the necessary conditions needed with special attention into the frequency of the input signal. Our findings are employed to elucidate if the generated charge due to piezoelectric effect overcomes the novel physical phenomena reported named "charge scrapping" arising from surface charge screening removal. Our data shows that piezoelectricity is the major contribution into the collected charge. Our findings are proved by scanning a Periodically Poled Lithium Niobate (PPLN) and a Lead Zirconate Titanate (PZT) ferroelectric material. We find that the charge collected is one order of magnitude higher for the case of PZT than for PPLN, while the screening charge density is lower in the PZT, compared with the PPLN. We then re-analyzed the data available for Charge Gradient Microscopy, specifically pointing out the conditions in which the experimental part was performed. We show that a transimpedance amplifier BW cannot be calculated using GBWP approximation, while working at higher frequencies than allowed increases the charge gain of the TIA. We prove that for CGM experiments, the images are recorded in the region in which the amplifier behaves as a charge amplifier, with a distinct transfer function. More over, the effect of feedback path delay is visible in the current profiles, as a constant frequency oscillation, while an exponential decayment, compatible with a "ringing" artifact. With this manuscript, we prove that piezoelectricity is the major contribution into the collected charge at the nanoscale.

ACKNOLEDGMENTS

I acknowledge comments and discussions from Marti Gich, Adrian Carretero Genevrier, Teresa Puig and Xavier Obradors. I acknowledge financial support from NFFA-Europe under the EU H2020 framework programme for research and innovation under grant agreement n. 654360. ICMAB acknowledges financial support from the Spanish Ministry of Economy and Competitiveness, through the "Severo Ochoa" Programme for Centres of Excellence in R&D (SEV- 2015-0496).